\documentstyle[prl,aps]{revtex}
\input epsf.tex

\newcommand{\nc}{\newcommand}
\nc{\beq}{\begin{equation}}
\nc{\eeq}{\end{equation}}
\nc{\beqa}{\begin{eqnarray}}
\nc{\eeqa}{\end{eqnarray}}
\nc{\lra}{\leftrightarrow}

\nc{\sss}{\scriptscriptstyle}
{\nc{\lsim}{\mbox{\raisebox{-.6ex}{~$\stackrel{<}{\sim}$~}}}
{\nc{\gsim}{\mbox{\raisebox{-.6ex}{~$\stackrel{>}{\sim}$~}}}

\def\Mv{M_{\sss V}}

\begin{document}
\twocolumn[\hsize\textwidth\columnwidth\hsize\csname@twocolumnfalse%
\endcsname

\draft

\title{Inflation from Extra Dimensions}

\author{James M.~Cline}

\address{Physics Department, McGill University,
3600 University Street, Montr\'eal, Qu\'ebec, Canada H3A 2T8}

\maketitle

\begin{abstract} The radial mode of $n$ extra compact dimensions (the
radion, $b$) can cause inflation in theories where the fundamental
gravity scale, $M$, is smaller than the Planck scale $M_P$.  For radion
potentials $V(b)$ with a simple polynomial form, to get the observed
density perturbations, the energy scale of $V(b)$ must greatly exceed
$M\sim 1$ TeV: $V(b)^{1/4} \equiv \Mv \sim 10^{-4} M_{P}$.  This gives
a large radion mass and reheat temperature $\sim 10^{9}$ GeV, thus
avoiding the moduli problem.  Such a value of $\Mv$ can be consistent
with the classical treatment if the new dimensions started sufficiently
small.  A new possibility is that $b$ approaches its stable value from
above during inflation.  The same conclusions about $\Mv$ may hold even
if inflation is driven by matter fields rather than by the radion.

\end{abstract}

\pacs{PACS: 98.80.Cq \hfill McGill 99-17}
]

The realization that compactified extra dimensions might have radii as
large as 1 mm without contradicting any current experiments \cite{ADD},
and that this could explain why the Higgs boson mass, $m_h$, is much
less than the Planck scale, has created a major new direction of research
in particle physics.  As long as the Standard Model particles are
somehow stuck on our three-dimensional subspace (brane) of the $3+n$
total spatial dimensions, the effects of the new large dimensions would
have so far escaped our notice.  In this picture, the Planck scale,
$M_P$, arises as a by-product of a much smaller fundamental scale, $M$,
and the radii of the compact dimensions, $b_0$, through the relation
\cite{norm}
\beq
\label{planck}
	M_P = 4 \sqrt{\pi} M (M b_0)^{n/2}.
\eeq
If $M$ is at the TeV
scale, then the problem of why $m_h$ is much smaller than $M_P$ is
superseded by explaining why $b_0$ is much larger than $M^{-1}$.
With large enough $n$ this is a less severe tuning problem than the
original hierarchy problem.

The major new effects for phenomenology come from gravitons and their
Kaluza-Klein excitations, which by (n+4)-dimensional general covariance
must be allowed to propagate in the extra dimensions, known as the
bulk.  Because their couplings to Standard Model particles are
suppressed by $M_P$, these effects can be kept duly small, yet near the
threshold for observation.  In addition to the gravitons, there is a
field, $b(t)$, associated with the variable size of the compact
dimensions, called the radion.  It can be thought of as the scale
factor for the compact dimensions, where the spacetime metric
has the Friedmann-Robertson-Walker-like form
$ g_{\mu\nu} = {\rm diag} (-1,\; a^2,\; a^2,\;
	a^2,\; b^2,\; \dots,\; b^2 ), $
provided that the compact space and the brane are homogeneous and
isotropic.  At late times, $b(t)$ must approach its asymptotic value
$b_0$.

When considering inflation in such theories, it is necessary to allow
for the evolution of $b(t)$ as well as the usual scale factor $a(t)$
which describes the growth of the large dimensions.  In fact, the
radion can economically play the role of the inflaton.
Ref.\ \cite{ADKM} has explored some of the possibilities, giving a
picture in which $b\ll b_0$ at the onset of inflation, grows slowly
during inflation, and attains its ultimate size only long after the end
of inflation.  The kind of radion potential which would give this kind
of behavior is rather complicated: for $b\sim b_i\ll b_0$, where $b_i$
is the value of $b$ during inflation, $V(b) \sim b^{2n}$; for $b_i \ll
b \ll b_0$, $V(b) \sim b^{-p}$, with $p>0$; and for $b\gg b_0$ one
expects that $V(b)\sim b^n$, since this is how the contribution from
the bulk cosmological constant scales with $b$.  The complications
arise from the requiring the right magnitude of density perturbations
without having to introduce any energy scales which are much greater
than the new gravity scale $M$.

In this letter we take a different attitude; we assume that the radion
potential has a relatively simple form, and then ask, what are the
consequences?  A major one is that, in order to obtain large enough
density perturbations, $V(b)$ must be much greater than $(1$ TeV)$^4$
during inflation.  While presenting a problem of naturalness, this
condition at the same time solves the moduli problem of the radion.  In
this scenario the radion stabilizes immediately following inflation.
There is also a new possibility: $b$ can overshoot its stable value and
approach $b_0$ from above during inflation.  We will present numerical
solutions of the evolution equations to illustrate these outcomes for
specific choices of the radion potential.  We also consider
conventional inflation driven by matter fields after the radion has
stabilized.  Interestingly, a similar restriction on $V(b)$ arises as
in the case when the radion is the inflaton, to avoid runaway inflation
of the compact dimensions.

We start by recapitulating the equations of motion and the consistency
condition for $a$ and $b$, derived in \cite{ADKM,ADM}. These can
be expressed in terms of Hubble parameters $H_a = \dot a/a$ and
$H_b = \dot b/b$,
\beqa
\label{abeqs}
&&	{\ddot a\over a} + 2 H_a^2 + n H_a H_b = {C_n\over b^{n}}
	\left( \left( b{d\over db} - (n-2)\right)V + 
	S_a \right) \nonumber\\
&&	{\ddot b\over b} + (n-1) H_b^2 + 3 H_a H_b = {C_n\over b^{n}}
	\left( \left(4 - 2{b\over n}{d\over db}\right)V 
	+S_b\right) \nonumber\\
&&	3 H_a^2 + \frac12 n(n-1) H_b^2 + 3 n H_a H_b =  (n+2){C_n\over
	b^{n}}(V + \rho)
\eeqa
where $C_n = (2(n+2) M^{n+2})^{-1}$, and $V(b)$ is a potential whose
minimum $b_0$ must be consistent with eq.~(\ref{planck}), and such that
$V(b_0) \cong 0$ so that there is no cosmological constant at the end
of inflation.  If there is significant pressure and energy density from
matter fields, say a scalar $\phi$ with potential $V_\phi$, then $S_a =
\rho + (n-1)p$, $S_b = \rho - 3 p$, $\rho = \frac12\dot\phi^2 + V_\phi$
and $p = \frac12\dot\phi^2 - V_\phi$, with the equation of motion
$\ddot\phi + 3 H_a \dot\phi + V'_\phi = 0.$ We shall at first assume
that $S_{a,b}$ are negligible compared to $V(b)$, however.

To get inflation of $a$ from $V$, it is necessary that $b$ starts out
away from its minimum, $b_i\neq b_0$, and rolls slowly toward $b_0$.
It is clear from eqs.\ (\ref{abeqs}) that the slow-roll condition is
$(4 - 2{b\over n}{d\over db})V \ll ( b{d\over db} - n+2 )V$
\cite{ADKM}.  This is only satisfied for a range of $b$ if $V$ has the
leading behavior $V(b) \sim b^{2n}$ during inflation.  For this work we
will consider a potential of the form
\beq
\label{Vbeq2}
	V(b) = \Mv^4\left( \hat b^{\alpha} - 
		\left( {\alpha-\gamma\over \beta-\gamma}\right)
		\hat b^\beta +
		 \left( {\alpha-\beta\over \beta-\gamma}\right)
                \hat b^\gamma \right)
\eeq
where $\hat b = b/b_0$.  It has the required properties that $V=V'=0$
at $b=b_0$; it is thus the simplest viable form.  There are two
possibilities for getting inflation:  either $b_i < b_0$, and $ \alpha
= 2n < \beta < \gamma$, or $b_i > b_0$ and $\alpha = 2n> \beta>
\gamma$.  The shape of $V$ in the two cases is illustrated in figure
1.  The initial condition where $b_i<b_0$ was investigated in
ref.\ \cite{ADKM}; here we will consider both possibilities.  Notice
that in either case, the middle term in $V$ is what drives $b$ during
inflation, since $(2 - {b\over n}\, {d\over db})$ must annihilate the
dominant $b^{2n}$ term in $V$ in order for $b$ to roll slowly.

\centerline{\epsfxsize=3.5in\epsfbox{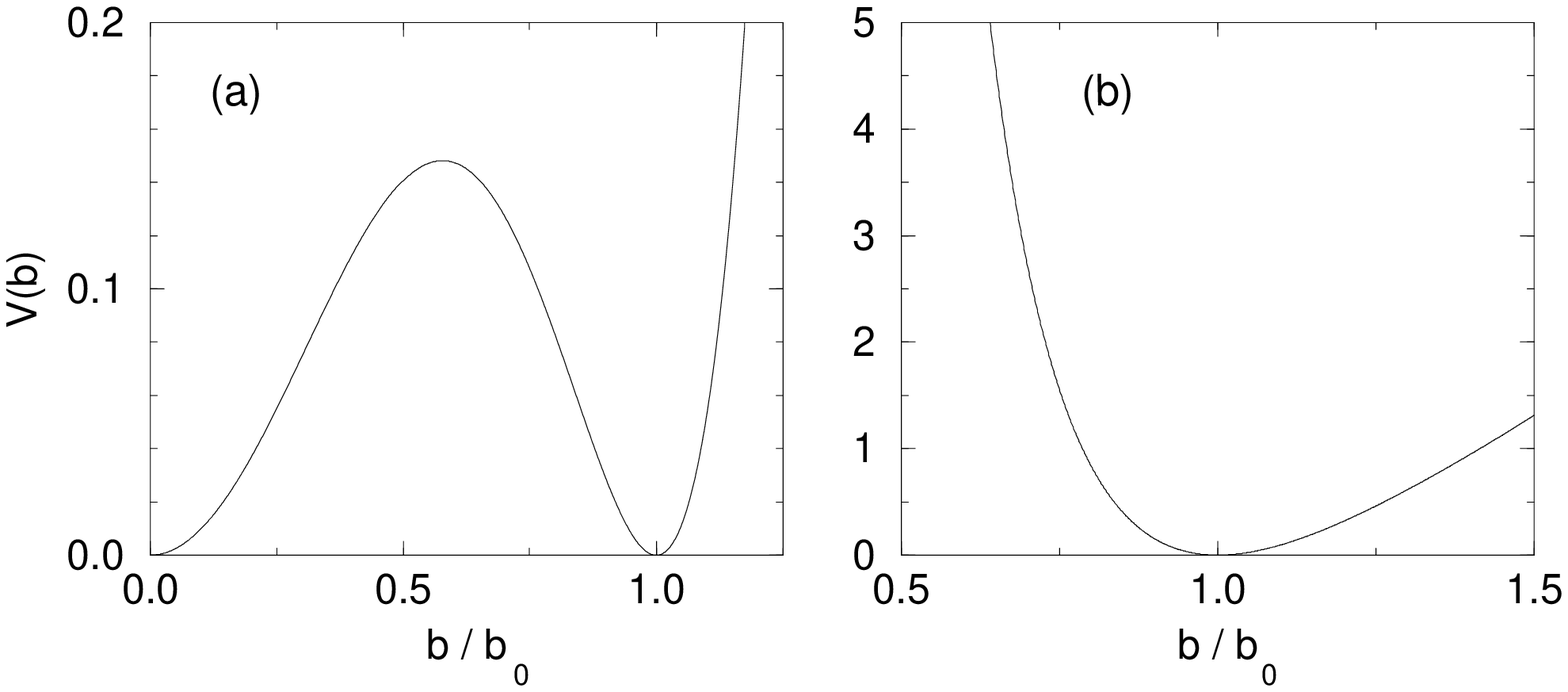}}
\vspace{-0.05in}
\noindent {\small 
Figure 1: radion potential for the case when
(a) $b_i < b_0$, so $V\sim b^{2n}$ at small $b$, and (b) $b_i>b_0$,
so $V\sim b^{2n}$ at large $b$.}
\vspace{+0.1in}

To numerically investigate the evolution of the scale factors and the
density perturbations produced during inflation, and also for
analytical insight, it is sometimes helpful to go to a dimensionless,
conformal-like time variable $\tau$, defined by ${db/dt} =
f(b) \; {db/d\tau}$.   
During inflation, it is convenient to choose 
$	f^2 = C_n b^{-n}  ( b {d\over db} - n+2 ) V
	\approx (n+2) C_n b^{-n} V,$
while letting $f$ become constant at the end of inflation.
Then the scale factor has the simple time dependence $a(t) \cong
e^{\tau/\sqrt{3}}$ during inflation.

The amount of inflation is determined mainly by $b_i/b_0$, 
the ratio of $b$'s initial and final values.
We will first discuss the case where $b_i < b_0$.
By approximately solving the radion equation of motion when $\ddot b\ll
H_b \ll H_a$, one finds the conformal time dependence of $b$ to be 
\beq
\label{bsoln}
 \tau = C_b \left((b_0/b_i)^{\beta-2n} - (b_0/b)^{\beta-2n}\right)
%\left. {\sqrt{3}n(n+2)(\beta-\gamma)\over
%	2(2n-\gamma)(\beta-2n)^2}  \left({b_0\over
%	b}\right)^{\beta-2n} \right|^{b_i}_{b(\tau)};
\eeq 
with $C_b = [\sqrt{3} n (n+2) (\gamma-\beta)] / [ 2 (\gamma-2n)
(\beta-2n)^2]$.  Thus the duration of inflation in conformal time goes
like $(b_0/b_i)^{\beta-2n}$.  Typically the conservative minimum of
$\sim70$ $e$-foldings of inflation (hence $\Delta\tau\approx 120$) can
be achieved starting with modest initial conditions, {\it e.g.,}
$b_i/b_0\le 0.07$, as illustrated in figure 2.  But we must also insure
that the observed magnitude of density perturbations is generated. At
COBE (Cosmic Background Explorer) scales, $\delta\rho/\rho \approx
1\times 10^{-5}$.  In the present theory, it is straightforward to show
that \cite{ADKM}
\beq 
\label{dp}
	{\delta \rho\over \rho} = {5\over 12\pi \sqrt{2n(n-1)}}\;
	{H_a^2 \over (bM)^{n/2} M H_b}\,,
\eeq
by relating the radion, $b$, to a canonically normalized inflaton
field.  We can derive a simple, accurate approximation for
$\delta\rho/\rho$ during inflation by solving for $H_a$ and $H_b$
during the slow-roll regime, when $\ddot b/b\ll H_b^2 \ll \ddot
a/b\cong H_a^2$. Recalling that $V(b)\sim \Mv^4$, one obtains, at an
epoch when $b=b_*$,
\vspace{-0.05in}
\beq
        {\delta \rho/ \rho}\,(b_*) = C_\rho \left({\Mv/
	M_P}\right)^2 \left({b_0/ b_*}\right)^{\beta-2n},
%\vspace{-0.05in}
\eeq
where $C_\rho = [5\sqrt{n} (n+2) (\beta-\gamma)]/[
	3\sqrt{3(n-1)} (2n-\beta) (2n-\gamma)]$.
The most natural value for $\Mv$ is the gravity scale $M$.  If $M$ is
only 1 TeV, as desired for solving the Higgs mass hierarchy problem,
the factor $(M/M_P)^2\sim 10^{-32}$ suppresses $\delta\rho/ \rho$
enormously. To compensate, one might be tempted to take the initial value 
of $b/b_0$ to be of order ${b_i / b_0} \sim (100\,
{\Mv/M_P})^{2/(\beta-2n)}.$

However, very small values of $b_i$ cause inflation to last much
longer, and the perturbations with large $\delta\rho/\rho$ were
produced so early that they are still beyond the present horizon.  COBE
perturbations, on the other hand, had a wavelength of $\lambda_d =
7\times 10^5$ ly at the time of photon decoupling (when $T = T_d =
0.25$ eV), and were produced 

\vspace{0.1in}
\centerline{\epsfysize=2.0in\epsfbox{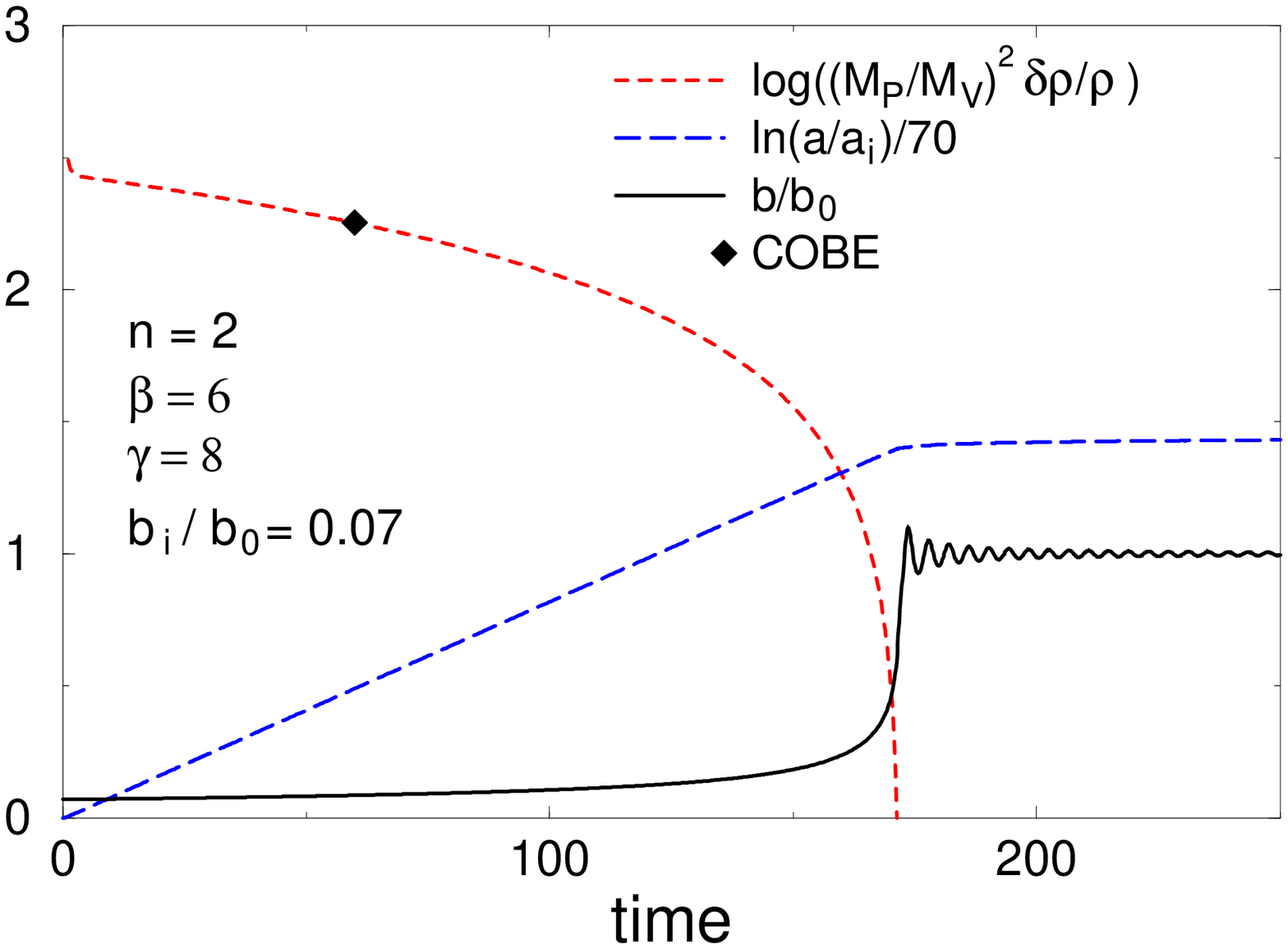}}

\noindent {\small \setlength{\baselineskip}{0pt}
Figure 2: $\log_{10}((M_P/\Mv)^2 \delta\rho/\rho$,
$\ln(a(t)/a_i)/70$ and $b(t)/b_0$, and
as a function of conformal time, as described in the text. 
The epoch at which COBE fluctuations were produced is marked by the
diamond.}
%The time coordinate is expanded by a
%factor of 5 after inflation to better show the radion oscillations.}

\noindent
$N$ $e$-foldings before the end of
inflation, with $N$ given by
\beq
\label{Nefolds}
	N \approx \ln \left({\lambda_d\; H_a}\,{T_d\over T_{rh}}\right)
	\sim \left(24-{14 n\over \beta-2n}\right)\ln 10.
\eeq
The second equality was obtained by assuming 
${b_i / b_0} \sim (100\, {\Mv/M_P})^{2/(\beta-2n)}$, $\Mv =
M = 1$ TeV, and a reheat temperature of 100 GeV; it shows that $N$ is at
most of order 10 under these assumptions.  Thus the relevant
perturbations were produced near the end of inflation.  They are not
enhanced by making $b_i$ smaller; their size is determined by the value
$b=b_*$ near the end of inflation, which is relatively close to $b_0$.

From the above results, the only way to get a large enough
$\delta\rho/\rho$ with our radion potential is to give it an energy
scale $\Mv\cong 3\times 10^{15}$ Gev, which is much greater than the
desired low gravity scale.  It is difficult to explain why $\Mv$ should
be so large if $M \sim 1$ TeV.  On the other hand, there must be a
large mass scale hidden in $V(b)$ in any case to obtain $b_0 \gg
M^{-1}$. Moreover, a large scale for $V(b)$ solves a second pernicious
problem: that of reheating \cite{ADD,CMT}.  The radion mass goes like
$m_b \sim \Mv^2/M_P \sim 10^{12}$ GeV, so it is no longer a modulus
which oscillates practically forever without decaying, as in the case
when $\Mv = 1$ TeV.  It decays quickly ($\Gamma_b \sim g_*
m_b^3/M_P^2$) into all $g_*$ species of lighter particles, since it
couples to the trace of the stress-energy tensor.

The conventional theory of reheating then gives a reheat temperature of
$T_{rh} \sim 10^{-1}\sqrt{\Gamma_b M_P}\sim 10^{9}$ GeV \cite{Linde}
from the decay of the radion condensate.  (We verified that the
resulting energy density is less than $V(b)$ at the end of inflation.)
Although such a large $T_{rh}$ will bring bulk gravitons into thermal
equilibrium, they decay with a rate $\Gamma\sim g_* T_{rh}^3/ M_P^2$,
thus disappearing well before nucleosynthesis \cite{BD}, and even
before baryogenesis; they decay in equilibrium. Although some gravitons
are produced with lower energy $E$, hence longer lifetimes, their
numbers are suppressed by the factor $(E/T_{rh})^n$ coming from the
production cross section.  Those with $E<10$ TeV, which would survive
to the era of nucleosynthesis and destroy light elements during their
decays, are therefore diluted by a factor of $\sim10^{5n}$ relative to
photons.

Once the reheating temperature is known, eq.\ (\ref{Nefolds}) can be
evaluated again to show that the COBE fluctuations were actually
produced at $N = 73 + n\ln(b_*/b_0)$ $e$-foldings before inflation.
This implicitly determines the value of $b_*$ since $N = (\Delta\tau -
\tau_*)/\sqrt{3}$, where $\Delta\tau$ is the duration of inflation, and
$b$ depends on $\tau$ according to eq.\ (\ref{bsoln}).  Numerically
solving in the example of figure 2 shows that the COBE perturbations
were produced at $\tau_* = 60$, $b_*/b_0=0.09$.  The perturbations are
rather flat in this region, with a spectral index of $n_\rho - 1 = 2
d\ln(\delta\rho/\rho)/d\ln(a) = -0.015$.  This is well within the
present observational limits, $|n_\rho - 1| < 0.2.$

One of the difficulties of making $\Mv\gg M$ is the validity of the
low-energy effective action which gives the equations of motion
(\ref{abeqs}) \cite{ADKM}.  One expects that regions of homogeneity,
where the spatial dependence of the metric can be neglected, will have
an extent of only $\Mv^{-1}$.  If this is much smaller than the size of
the compact dimensions, then the assumption of homogeneity within the
compact space is unjustified.  However with the potential
(\ref{Vbeq2}), it is possible to start with $V(b_i) \ll \Mv^4$ since
$b_i$ might be much smaller than $b_0$.  In such a region it
is natural for the spatial fluctuations in the geometry to have a
longer wavelength than $b_i$.  The condition for having $V(b_i)^{1/4} <
b_i^{-1}$ can be written as
\beq
\label{infcond}
	b_i/b_0 < (M_P/\Mv)^{2/(n+2)} (M/M_P)^{2/n}\, .
\eeq
If $M = 1$ TeV, then the least fine-tuned case of $n=7$, assuming an
upper limit of 11 dimensions, gives $b_i/b_0 < 10^{-4}$.  In the spirit
of chaotic inflation, this is not unnaturally small.  As long as there
are some regions in the initial universe satisfying (\ref{infcond}),
they will start to inflate, and continue for many
($\sim10^{4(\beta-2n)}$) $e$-foldings.  If other regions fail to
inflate because of their inhomogeneities, this need not concern us.
Moreover, if $M$ is of order $\Mv$, eq.\ (\ref{infcond}) becomes
$b_i/b_0 \lsim 10^{-12/(n(n+2))}$, which needs no fine-tuning at all if
$n>3$.

The above discussion was for $b_i < b_0$.  What happens if the compact
space starts out larger than its stable value?  As long as the exponent
$\beta$ is less than $2n$ in the potential (\ref{Vbeq2}) and $\gamma <
\beta$, the story is quite similar to the case of $b_i < b_0$.  One can
obtain enough inflation if $b_i/b_0\gsim 6$, and the estimate of the
density fluctuations again requires $\Mv$ to be near the GUT scale.
However the initial conditions now appear to be quite fine-tuned:  the
length scale provided by the potential, $\Mv^{-1}$, is many orders of
magnitude smaller than the size of the compact space, so one must
wonder how the latter came to be so smooth at the beginning of
inflation.  In fact, it is possible to start with $b\ll b_0$, as
before, but the radion picks up so much speed that it overshoots $b_0$
and attains a large value which marks the onset of inflation.  This
situation is illustrated in figure 3.  The spectral index is again
$n_\rho -1 = -0.015$ in this example.

\vspace{0.1in}
\centerline{\epsfysize=2.0in\epsfbox{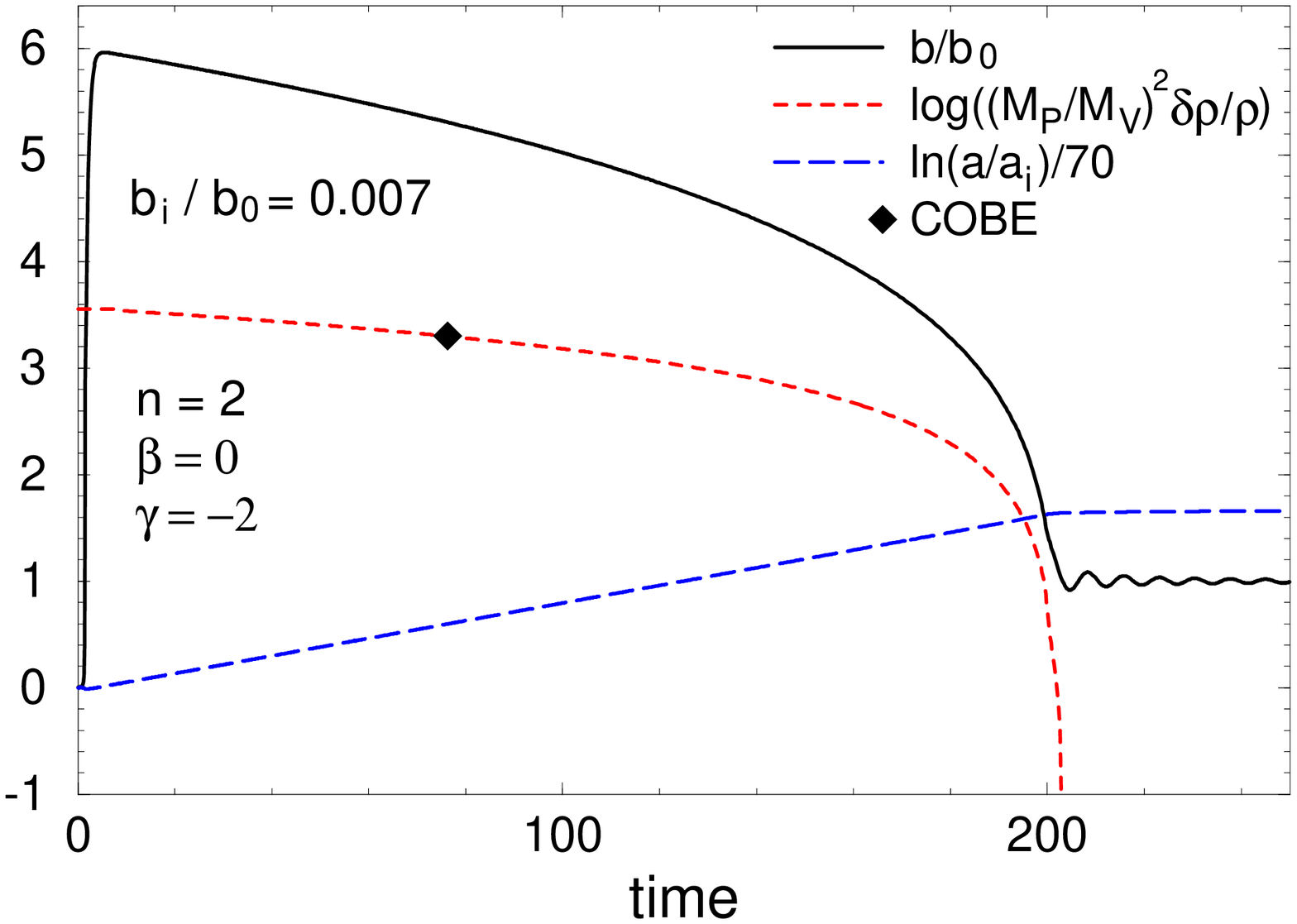}}
\vspace{-0.0in}
\noindent {\small Figure 3: $b(t)/b_0$, 
$\log_{10}((M_P/\Mv)^2 \delta\rho/\rho$ and 
 $\ln(a(t)/a_i)/70$ in an example where
inflation occurs while $b > b_0$.}

Unfortunately the most natural form for the radion potential is $ V(b)
\sim Ab^n - Bb^{n-2} + Cb^{n-2p}$ \cite{ADM,BDN}, which does not give
inflation.  For this reason it may be more convincing to drive inflation
with conventional matter fields \cite{KL} or extra branes
\cite{DT} after the radion has stabilized.  An interesting implication
of the extra dimensions is that they can be destabilized by a
conventional inflaton, and begin inflating themselves.  As one might
expect, whether this occurs depends upon how stiff the radion potential
is compared to the size of the perturbing matter potential, which we
shall refer to as $V_\phi$.  Thus one obtains a constraint on $V(b)$
from the properties of the inflaton potential.  (A different argument
leading to a similar constraint was given in refs.\ \cite{BD}
and \cite{Lyth}.)

To see how such a constraint comes about, 
parametrize the radion by $b = b_0(1+\epsilon)$ and 
suppose that the potential has the form $V \sim
\Mv^4 \epsilon^2$ in the vicinity of $b_0$.  Consider an inflationary
phase where the pressure and energy density are dominated by the
potential $V_\phi$ of a matter field $\phi$ that lives on the brane.
(The same conclusions also follow if $\phi$ inhabits the bulk.) The
source terms in eq.\ (\ref{abeqs}) are $S_a = (2-n)V_\phi$ and $S_b = 4
V_\phi$.  A nonzero $S_b$ will induce a shift in the radion as it tries
to minimize the full source term for the $b$ equation of motion,
\beq
\label{sourceb}
	-n^{-1} \Mv^4\, \epsilon \,( 1-(n-1)\, \epsilon )
	+ V_\phi \cong 0 \,.
\eeq
If $V_\phi\ll \Mv^4$ we can work to linear order in $\epsilon$.  The
source term for the $a$ equation then becomes $(n+2) C_n b^{-n}
V_\phi$, in accordance with the consistency condition, third of
eqs.\ (\ref{abeqs}).  However if $V_\phi$ becomes too large, the
quadratic term in $\epsilon$ becomes important, and it is no longer
possible to satisfy eq.\ (\ref{sourceb}).  At this point $b$ cannot
resist inflation.  This situation must be avoided because, once $b$
starts inflating, it is a runaway process; $b$ inflates forever, as
shown in figure 4.

If $b$ remains close to $b_0$, eqs.\ (\ref{abeqs}) reduce to the usual
Friedmann equation, $H_a^2 = 8\pi \rho/3M_P^2$.  In chaotic inflation
with a potential of the form $V_\phi = M_\phi^{4-p}\phi^p$, it
is straightforward to show that the relation between $\phi$ and $a$ is
$\phi^2 = \phi_i^2 - (p M_P^2/4\pi)\ln(a/a_i)$, so that initially 
$\phi_i \sim$ (several $M_p)$ to get 70 $e$-foldings of $a$.  The density 
perturbations during this epoch are given by 
$\delta\rho/\rho \sim (5/12\pi) H_a^2/\dot\phi \sim 10^3 \sqrt{V_\phi}
M_P^{-2} \sim 10^{-5}$.  We thus obtain \vspace{-0.1in}
\beq
	V_\phi \sim 10^{-16} M_P^4 \lsim \Mv^4
\eeq
which is quite similar to the bound on $\Mv$ when the radion is the
inflaton.  In hybrid inflation models \cite{Linde2} this bound is
softened to $\Mv > 10^{-1}M_P^{3/5}m^{2/5}$.

Finally let us ask: how problematic is it for $\Mv$ to be much larger
than the gravity scale, assuming it is near 1 TeV?  First, it should be
kept in mind that there is still no proposal for getting the required
$b^{2n}$ behavior of $V(b)$ (nor the subsequent $b^{-p}$ behavior in
the post-inflation expansion period envisioned in \cite{ADKM}).
Whatever mechanism that emerges to explain this might also make a large
energy scale more plausible.  Second, even in the most realistic model
for $V(b)$, scales larger than $M$ may be required to explain why $b_0
\gg M^{-1}$.  The term in $V(b)$ coming from $\sqrt{-g}R$ in the
Einstein-Hilbert action is necessarily of order $M^4 (Mb)^{n-2}$, so
that when $b\sim b_0$, $V\sim M^4 (M_P/M)^{2 - 4/n}$, which is much
larger than $M^4$ unless $n=2$.  Although not as severe a hierarchy
problem as the present one, it is qualitatively similar.  Third, the
$\Mv$ hierarchy problem disappears altogether if we allow $M$ to be the
GUT scale rather than 1 TeV.   While this value is less exciting for
current accelerator experiments, it may be nature's choice, and it
still presents new possibilities for the early universe.

I thank C.\ Burgess, N.\ Kaloper, G.\ Moore, R.\ Myers and
M.\ Paranjape for helpful observations and criticisms.

\vspace{0.1in}
\centerline{\epsfysize=2.2in\epsfbox{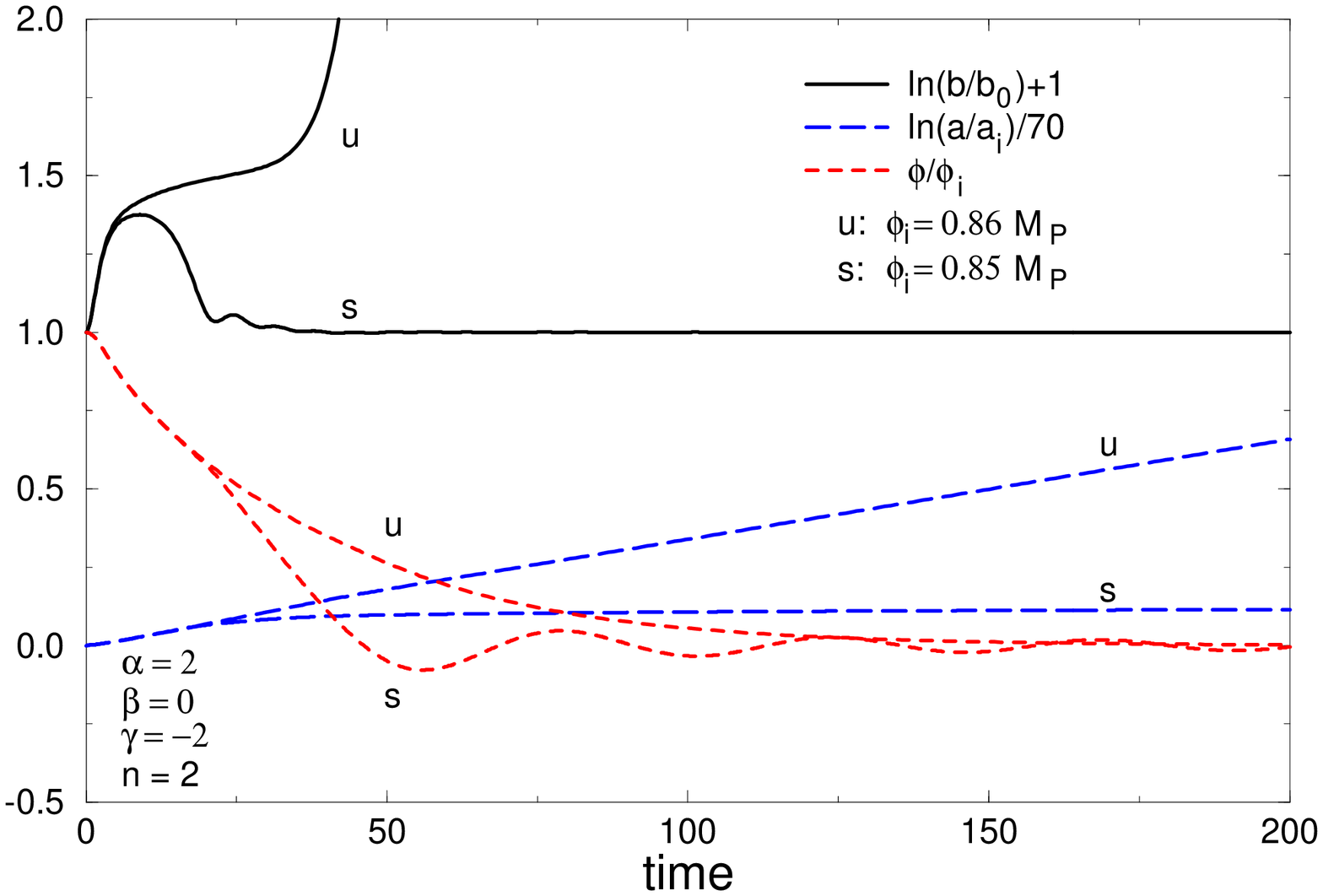}}
\vspace{-0.0in}
\noindent {\small Figure 4: $\ln(b(t)/b_0)+1$, 
$\ln(a(t)/a_i)/70$ and $\phi/\phi_i$ for initial conditions that
are stable (s) and unstable (u)
against eternal inflation of $b$.  $V_\phi$ is proportional to $\phi^2$
in this example.}

\vspace{-0.1in}

%\begin{thebibliography}{99}  

\end{document}